\documentclass[prb,letterpaper,twocolumn,groupedaddress,amsmath,amssymb,showpacs]{revtex4}

\usepackage{graphicx}
\usepackage{bm,mathrsfs}
\begin{document}

\title{Size-dependence of Strong-Coupling Between Nanomagnets and Photonic Cavities}
\author{\"O. O. Soykal}
\author{M. E. Flatt\'e}
\affiliation{
Optical Science and Technology Center and Department of Physics and Astronomy\\
University of Iowa, Iowa City, IA 52242}

\begin{abstract}
The coherent dynamics of a coupled photonic cavity and a nanomagnet is explored as a function of nanomagnet size. For sufficiently strong coupling eigenstates involving highly entangled photon and spin states are found, which can be combined to create coherent states. As the size of the nanomagnet increases its coupling to the photonic mode also monotonically increases, as well as the number of photon and spin states involved in the system's eigenstates.  For small nanomagnets the crystalline anisotropy of the magnet strongly localized the eigenstates in photon and spin number, quenching the potential for coherent states. For a sufficiently large nanomagnet the macrospin approximation breaks down and different domains of the nanomagnet may couple separately to the photonic mode. Thus the optimal nanomagnet size is just below the threshold for failure of the macrospin approximation.
\end{abstract}


\maketitle

\section{Introduction}

Coupling between an electromagnetic field and an electronic transition in matter, with coupling stronger than environmental dissipation, has permitted delicate electromagnetic control of electronic states. This control allows sensitive measurement of unknown environments, such as the extensive use of nuclear magnetic resonance\cite{Slichter1963} (NMR) as a diagnostic probe, as well as the manipulation of quantum information, such as in the demonstration of atomic teleportation\cite{Barrett2004}. Successful efforts in this area have tended to progress from the systems most weakly coupled to the environment (such as nuclei in NMR) to more dissipative systems (such as electron spin resonance in solids, first in insulators, later in metals and semiconductors).  However, even though the dissipation is stronger in solids, the coupling is also stronger, suggesting the potential for very rapid exchange of quantum states between light and matter.   Recently several examples of strong coupling between a single exciton and a single photon in a semiconductor have been demonstrated, through the mixing of the exciton and photon in photoluminescence\cite{Reithmaier2004,Yoshie2004}, through Rabi oscillations between exciton and photon\cite{Zrenner2002}, and through optically-induced spin rotation\cite{Gupta2001b,Berezovsky2008,Fu2008,Press2008} (spin AC Stark effect). 

Multiply-excited atomic systems coherently interacting with a photon mode exhibit additional unusual phenomena, such as superradiance\cite{Dicke1954}. Multiply-excited excitonic systems in solids suffer from decoherence due to homogeneous and inhomogeneous linewidths and long-range F\"orster coupling between different excitonic transitions. Mitigation of both homogeneous and inhomogeneous linewidths is possible by coupling the excitons to each other through the coulomb interaction (excitonic condensate\cite{Zimmermann1976}), or indirectly through the cavity mode, in order to form a polariton condensate\cite{Kasprzak2006,Balili2007}. Excitonic condensates associated with finite-energy excitons, however, are challenging to generate, and are not found at room temperature. Strong coupling in multiply-excited systems would therefore benefit from a robust, room-temperature, coherent electronic state whose coupling to the photonic mode can be made larger than its decoherence rate.

Ferromagnets are robust room-temperature many-body states that couple directly to light, although the magnetic dipole transitions associated with individual spins couple more weakly to photons than electric dipole transitions (by a factor of the fine structure constant\cite{Jackson1998b}).  Recently it has been pointed out\cite{Soykal2010} that the coherent excitation of the ground-state spin of a small ferromagnet (a nanomagnet) can be described by a coupling strength orders of magnitude stronger than that of a single excitonic transition. The locking of the large number of constituent spins by the exchange interaction into a macrospin causes an increase in coupling strength proportional to the square root of the number of exchange-locked spins. 

Here we expand on the description in Ref.~\onlinecite{Soykal2010}, treating with particular care the dependence on nanomagnet size of the coupling strength between the nanomagnet and the photonic cavity. We find that, for a specific magnetic material, the coupling strength increases according to the square root of the volume of the nanomagnet (corresponding to the square root of the total nanomagnet spin) in the absence of any photons inside the cavity. However, when the system is driven in the superradiance regime, this coupling strength becomes proportional to the volume (or total spin) to the $3/2$ power. We provide estimates of the coupling strength for nanomagnets in a spherical cavity and compare with the effect of crystalline magnetic anisotropy. The coupling strengths found are large enough to establish eigenstates involving large numbers of entangled photons and spin orientation states. These states can be combined to generate coherent oscillations of the spin orientation and photon number.  However, for small nanomagnets the crystalline magnetic anisotropy greatly exceeds the nanomagnet-cavity coupling, quenching these coherent oscillations. For large nanomagnets the macrospin approximation, assumed here, fails and the multiple domains of the nanomagnet separately couple to the cavity. The effect of using plasmonic techniques to enhance the magnetic field associated with the photonic mode near the nanomagnet is described, which may lead to sub-microsecond oscillation times for coherent multiphoton oscillations in the cavity.

We begin by describing the nanomagnet-cavity system, quantizing the photons of the spherical cavity, and deriving the Hamiltonian of the system. We solve for the eigenstates of the coupled system by mapping the discrete system onto a continuum representation similar to a one-dimensional tight-binding model with a spatially-varying effective mass. Perturbations to the magnetic system such as magnetic anisotropy can be described as spatially-varying potentials for this one-dimensional tight-binding model. The time-evolution of coherent states is evaluated, and the source of dephasing discussed.

\section{Coupled Nanomagnet-Cavity Formalism}

\subsection{Nanomagnet Properties}

\begin{figure}[htp]
  \centering
  \includegraphics[width=6.0 cm]{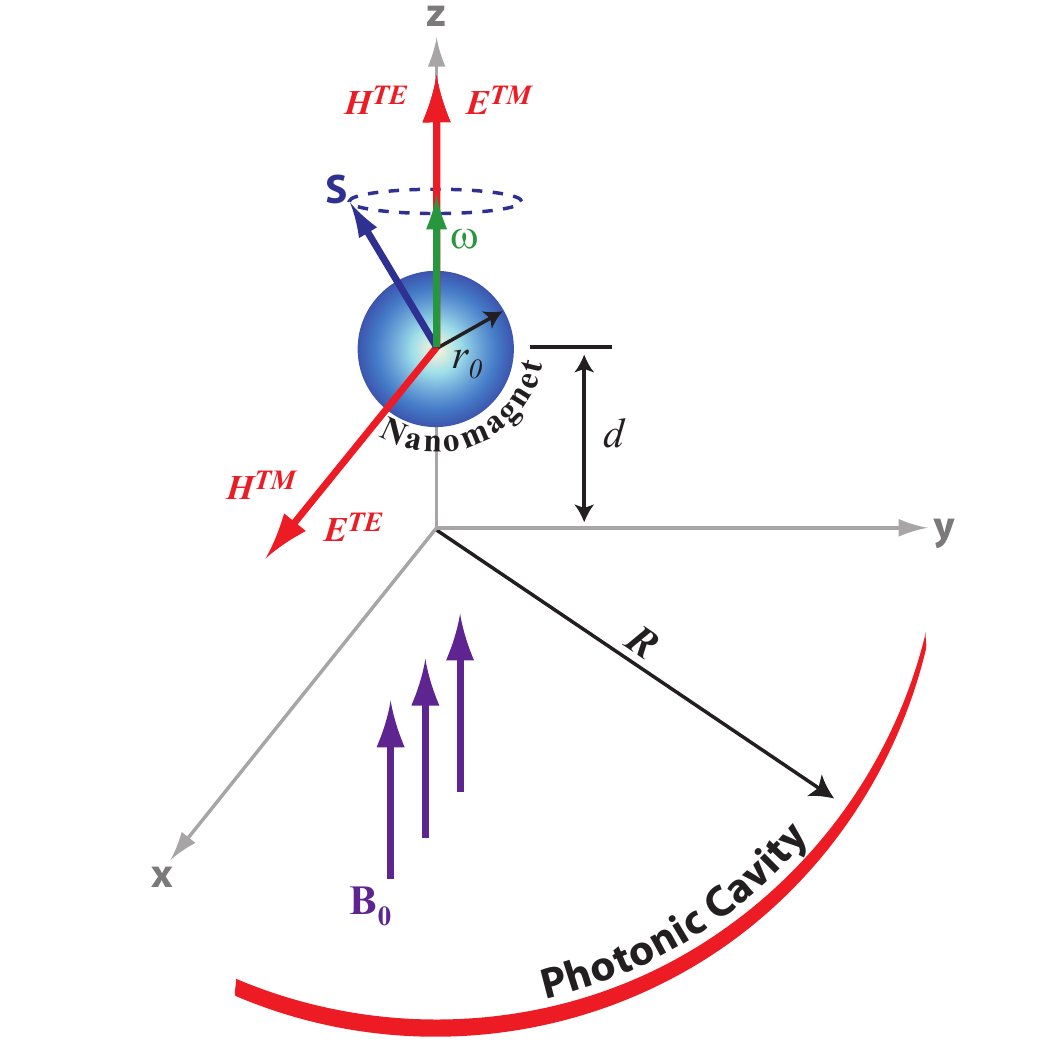}
  \caption{(color online) Schematic of the nanomagnet-cavity system with a spherical nanomagnet of radius $r_0$ placed at a distance of $d$ from the center of a spherical photonic cavity of radius $R$. The orientations of the electric $\bm{E}$ and magnetic field $\bm{H}$ at the nanomagnet site are shown for transverse magnetic (TM) and transverse electric (TE) modes of the photonic cavity. A uniform magnetic field, $\bm{B}_0$, applied along the $\mathbf{z}$-axis causes precession of the nanomagnet macrospin, $\bm{S}$, with frequency $\omega$, in resonance with the TM mode of the cavity.}\label{fig:schematic}
\end{figure}

As shown schematically in Fig.~\ref{fig:schematic}, the oscillator is a spherical nanomagnet with a radius $r_0$ possessing a very large exchange-locked spin $\bm{S}$ placed a distance $d$ from the center of the cavity for more efficient coupling to the cavity mode. Precession of the nanomagnet macrospin at a frequency $\omega$ resonant with the cavity is achieved by applying a uniform magnetic field $\bm{B}_0$ along the $\mathbf{z}$-axis of the cavity. 

A nanomagnet acting as a macrospin, as seen experimentally in nanomagnet oscillators of roughly this size\cite{Sankey2006}, has a magnetization
\begin{equation}
\bm{M}=\bm{\mu}/V=-\frac{g_s\mu_B}{\hbar V}\bm{S}\Theta(r_0-|\bm{r}-\bm{d}|),\label{13}
\end{equation}
in terms of the collective spin operator $\bm{S}$ and the Heavyside step function $\Theta(x)$.  The magnetization in Eq.~(\ref{13}) depends on the spin density of the nanomagnet. The modal coupling (the coupling of the nanomagnet to the photonic mode) is the overlap of this magnetization with the cavity mode amplitude. For a nanomagnet that is small in size compared to the length scale of variations in the cavity mode strength, the coupling will be independent of the spin density and will only depend on the total spin.  It is possible, however, to enhance the modal coupling through mode design, such as is common to enhance the interaction between gain media and an optical cavity in semiconductor lasers\cite{Agrawal1993}. For example, an optical field is stronly enhanced near a sharp metal object (used in tip-enhanced spectroscopy\cite{Moskovits1985}); a similar approach here could be used to strongly enhance the strength of the nanomagnet-cavity coupling. 

\subsection{Quantized electromagnetic field in a spherical cavity with a nanomagnet}

The presence of the nanomagnet in the cavity, and its magnetization field, modifies the properties of the dynamic electromagnetic field in the cavity. The nanomagnet precesses in the static external magnetic field, yielding a temporally-oscillating magnetization characterized by the precession frequency $\omega$. Thus the nanomagnet behaves as an oscillating source in the Maxwell equations
\begin{equation}
\begin{array}{rcl rcl}
\bm{\nabla}\cdot \bm{H} & = & 0,\quad\quad & \bm{\nabla}\times \bm{E}-ik\sqrt{\mu_0/\epsilon_0}\bm{H} & = & 0,\\
\bm{\nabla}\cdot \bm{E} &= & 0,\quad\quad & \bm{\nabla}\times\bm{H} +ik\bm{E}/\sqrt{\mu_0/\epsilon_0} & = & \bm{\nabla}\times\bm{M}.
\end{array}\label{1}
\end{equation}
We introduce the time dependence of the fields ($e^{i\omega t}$) into $\bm{H}$, $\bm{E}$, and $\bm{M}$. This produces the following Helmholtz wave equations,
\begin{eqnarray}
\left(\nabla^2+k^2\right)\left(\bm{r}\cdot\bm{H}\right)&=& -i\bm{L}\cdot\left(\nabla\times\bm{M}\right),\nonumber\\
\left(\nabla^2+k^2\right)\left(\bm{r}\cdot\bm{E}\right)&=& Z_0k\bm{L}\cdot\bm{M}.\label{2}
\end{eqnarray}
From these the solutions of the transverse magnetic (TM) and electric modes (TE) can be obtained
\begin{widetext}
\begin{eqnarray}
\bm{H}&=&\sum_{l,m}\left[\alpha_{lm}^{(TM)}f_{l}(kr)\mathbf{Y}_{l,l,m}
(\theta ,\phi)\right.
\left.-\frac{i}{k}\alpha_{lm}^{(TE)}\bm{\nabla}\times g_{l}(kr)\mathbf{Y}_{l,l,m}(\theta ,\phi)\right],\nonumber\\
\bm{E}&=&Z_0\sum_{l,m}\left[\frac{i}{k}\alpha_{lm}^{(TM)}\bm{\nabla}
\times f_{l}(kr)\mathbf{Y}_{l,l,m}(\theta ,\phi)\right.
\left.+\alpha_{lm}^{(TE)}g_{l}(kr)\mathbf{Y}_{l,l,m}(\theta ,\phi)\right],
\label{4}
\end{eqnarray}
\end{widetext}
where the vector spherical harmonics $\mathbf{Y}_{l,l,m}$ are defined as $\bm{L}Y_{lm}(\theta,\phi)/\sqrt{l(l+1)}$, in terms of angular momentum operator of the field $\bm{L}$ and spherical harmonics $Y_{lm}$.\cite{Heitler1998} In the most general form, they are defined as 
\begin{equation}
Y_{j,l,m_j}=C\langle l,1;m_l,m|j,m_j\rangle Y_{lm_l}\mathbf{\hat{e}_m}
\end{equation}
 in terms of Glebsch-Gordan coefficients and helicity basis vectors $\mathbf{\hat{e}_m}$. The helicity basis vectors form a spherical tensor of rank $1$, i.e. 
 \begin{equation}
 \mathbf{\hat{e}_\pm}=\mp(\mathbf{\hat{x}}\pm i\mathbf{\hat{y}})/\sqrt{2},
 \end{equation}
where $\mathbf{\hat{e}_0}=\mathbf{\hat{z}}$. 
 
The  $f_l (kr)$ and $g_l (kr)$ appearing in Eq.~(\ref{4}) are the corresponding solutions for the radial part of each mode,  $A_l^{(1)}(kr)h_l^{(1)}(kr)+A_l^{(2)}(kr)h_l^{(2)}(kr)$, in terms of the spherical Hankel functions. The coefficients $\alpha_{lm}^{(TM)}$ and $\alpha_{lm}^{(TE)}$, which specify the amounts of transverse magnetic and transverse electric multipole $(l,m)$ field strengths, are 
\begin{eqnarray}
\alpha_{lm}^{(TM)}&=&\frac{ik^3}{\sqrt{l(l+1)}}\int j_l (kr^\prime)Y_{lm}^* (\theta^\prime,\phi^\prime)\bm{L\cdot M}d^3 r^\prime ,\label{5}\\
\alpha_{lm}^{(TE)}&=&\frac{-k^2}{\sqrt{l(l+1)}}\int j_l (kr^\prime) Y_{lm}^* (\theta^\prime,\phi^\prime)\bm{L}\cdot\left(\bm{\nabla}\times\bm{M}\right) d^3 r^\prime ,\nonumber
\end{eqnarray}
where the volume integration is carried over the local sources. 

The radiation of the cavity field is due to the harmonically oscillating components of the nanomagnet magnetization in the x-y plane orthogonal to the radial direction, namely $\bm{M}_{x,y}$. Because of this specific symmetry of the cavity-nanomagnet system, the multipole field strength coefficients $\alpha_{lm}^{(TE)}$ for TE mode will simply vanish due to the relation 
\begin{equation}
\bm{L}\cdot\left(\bm{\nabla}\times\bm{M}\right)= i\nabla^2\left(\bm{r}\cdot\bm{M}\right)-\frac{i}{r}\frac{\partial}{\partial r} \left(r^2 \bm{\nabla}\cdot\bm{M}\right),
\end{equation}
 which holds for  any well-behaved vector field. Therefore, the TM mode will be the only non-vanishing mode to be considered in our interaction Hamiltonian. The condition $H_\perp=0$ is trivially satisfied by $\bm{r}\cdot\mathbf{u}_{lm}=0$ at the cavity walls, whereas the condition $E_\parallel=0$ gives
\begin{equation}
\left[\bm{r}\times\left(\bm{\nabla}\times f_l (kr)\mathbf{Y}_{l,l,m}\right)\right]|_{r=R}=-\partial_r\left(rf_l (kr)\right)\mathbf{Y}_{l,l,m}=0.\label{6}
\end{equation}

For waves that are finite at the origin, the suitable choice of $f_l(kr)$
is the spherical Bessel function of first kind, $j_l(kr)$. Hence, the normalization integral of the basis functions $\bm{u}_{lm}=j_l(kr)\mathbf{Y}_{l,l,m}$, modified for $x_{l\gamma}$ (corresponding to the zeros of $|rj_l (kr)|^\prime$),
\begin{eqnarray}
&&\int\bm{u}_{lm}^*\bm{u}_{l^\prime m^\prime}d^3 r = \int j_{l}(kr)j_{l^\prime}(kr)\mathbf{Y}_{l,l,m}^*\mathbf{Y}_{l^\prime,l^\prime, m^\prime}r^2 dr d\Omega\nonumber\\
&&\qquad\qquad=\frac{R^3}{2}\left(1-\frac{l(l+1)}{x_{l\gamma}^2}\right)|j_{l}(x_{l\gamma})|^2 \delta_{l l^\prime}\delta{m m^\prime},\label{7}
\end{eqnarray}
yields to the following mapping of the multipole strength coefficients onto cavity photon creation and annihilation operators:
\begin{eqnarray}
\alpha_{lm}^{(TM)}&\longmapsto&\frac{2}{|j_{l}(x_{l\gamma})|} \left[1-\frac{l(l+1)}{x_{l\gamma}^2}\right]^{-1/2}\sqrt{\frac{\hbar \omega_{l\gamma}}{\mu_0 R^3}}a_{lm}^{(TM)},\nonumber\\
\alpha_{lm}^{*(TM)}&\longmapsto&\frac{2}{|j_{l}(x_{l\gamma})|} \left[1-\frac{l(l+1)}{x_{l\gamma}^2}\right]^{-1/2} \sqrt{\frac{\hbar\omega_{l\gamma}}{\mu_0 R^3}}a_{lm}^{\dagger (TM)},\nonumber
\end{eqnarray}
which satisfy the appropriate Weyl-Heisenberg commutation relations, $[a_{lm},a_{l^\prime m^\prime}^\dagger]=\delta_{ll^\prime}\delta_{mm^\prime}$.

Therefore, the second quantized form of the magnetic field for the cavity TM mode becomes
\begin{widetext}
\begin{equation}
\bm{H}^{(TM)}=\sum_{l,m}\frac{1}{|j_l (x_{l\gamma})|}\left[1-\frac{l(l+1)}{x_{l\gamma}^2}\right]^{-1/2} \sqrt{\frac{\hbar\omega_{l\gamma}}{\mu_0 R^3}}\left(a_{lm}^{\dagger (TM)} \mathbf{u}_{lm}^* +a_{lm}^{(TM)}\mathbf{u}_{lm}\right).\label{8}
\end{equation}
\end{widetext}

The total Hamiltonian of the system incorporates the magnetic $\bm{H}$ and electric $\bm{E}$ fields of the cavity and the magnetization $\bm{M}$ of the nanomagnet\cite{Jackson1998},
\begin{equation}
\mathscr{H}=\frac{1}{2}\int\left(\mu_0|\bm{H}|^2+\epsilon_0|\bm{E}|^2+\mu_0
\left(\bm{H}\cdot\bm{M}\right)\right)d^3r.\label{9}
\end{equation}
The first two integrands on the right hand side of Eq.~(\ref{9}) correspond to the free field Hamiltonian, whereas the third integrand is the interaction Hamiltonian of the nanomagnet-cavity system,
\begin{equation}
\mathscr{H}_I=\sum_{l,m}\Gamma_{l}^{(TM)} a_{lm}^{(TM)}\int_{V_m}\bm{M}\cdot\bm{u}_{lm}\,d^3r + c.c.\label{10}
\end{equation}
with the coupling constant,
\begin{equation}
\Gamma_{l\gamma}^{(TM)}=\frac{1}{2|j_l (x_{l\gamma})|} \left[1-\frac{l(l+1)}{x_{l\gamma}^2}\right]^{-1/2} \sqrt{\frac{\hbar\omega_{l\gamma}\mu_0}{R^3}},\label{11}
\end{equation}
for a TM mode with angular momentum $l$. All components of the field are identically zero if $l=m=0$, a result associated with the absence of radiating monopoles. From Eq.~(\ref{5}) that dipole field strength coefficient ($l=1$) dominates over other multipoles, i.e. $\alpha_{1m}^{(TM)}\gg\alpha_{2m}^{(TM)}\gg\alpha_{3m}^{(TM)}\gg\dots$. The basis functions for the dominant dipole TM mode ($l=1$) are
\begin{eqnarray}
\mathbf{u}_{11}&=&\frac{1}{\sqrt{2}}j_1 (kr)\left(Y_{11} (\theta,\phi)\bm{\hat{e}}_{0}-Y_{10} (\theta,\phi)\bm{\hat{e}}_{+}\right),\nonumber\\
\mathbf{u}_{10}&=&\frac{1}{\sqrt{2}}j_1 (kr)\left(Y_{11} (\theta,\phi)\bm{\hat{e}}_{-}-Y_{1\bar{1}} (\theta,\phi)\bm{\hat{e}}_{+}\right),\label{12}\\
\mathbf{u}_{1\bar{1}}&=&\frac{-1}{\sqrt{2}}j_1 (kr)\left(Y_{1\bar{1}} (\theta,\phi)\bm{\hat{e}}_{0}-Y_{10} (\theta,\phi)\bm{\hat{e}}_{-}\right),\nonumber
\end{eqnarray}

\subsection{Coupling of the Nanomagnet to the Photonic Cavity}

To describe the coupling of the nanomagnet to the cavity, the spin operators of the nanomagnet should be written in the same helicity basis as the photonic field,
\begin{equation}
\bm{S}=\frac{1}{\sqrt{2}}(S_+\mathbf{\hat{e}_-}-S_-\mathbf{\hat{e}_+})+S_z\mathbf{\hat{e}}_0,\label{14}
\end{equation}
in terms of the nanomagnet spin raising and lowering operators 
\begin{equation}
S_{\pm}|l_s,m_s\rangle=\sqrt{(l_s\mp m_s)(l_s\pm m_s+1)}|l_s,m_s\pm 1\rangle.
\end{equation}
 Introduction of this total spin operator $\bm{S}$ and the basis functions of the spherical wave expansion (Eq.~(\ref{12})) into Eq.~(\ref{10}) yields a fully quantum treatment of the total Hamiltonian for the nanomagnet-cavity system,
\begin{eqnarray}
\mathscr{H}_\gamma&=&\hbar\omega_\gamma\left(a_\gamma^\dagger a_\gamma+\frac{1}{2}\right)+g\frac{\mu_B}{\hbar}B_0 S_z\nonumber\\
&&-g\mu_B\Gamma_\gamma
\left(a_\gamma S_+ + a_\gamma^\dagger S_-\right),\label{15}
\end{eqnarray}
in which the spin interacts only with a single photon mode $\gamma$. Modes of higher $\ell$ would be out of resonance because of the cavity quantization, and energy non-conserving terms with negative helicity have been dropped (relying on the rotating wave approximation \cite{Scully1997}). The nanomagnet-photon coupling constant $\Gamma_\gamma$ becomes
\begin{equation}
\Gamma_\gamma=\frac{j_1(kd)}{8\hbar|j_1(y_{1\gamma})|}\left[1-\frac{l(l+1)}{y_{1\gamma}^2}\right]^{-1/2}
\sqrt{\frac{3\hbar\omega_\gamma\mu_0}{\pi R^3}},\label{16}
\end{equation}
where the mode frequency $\omega_{\gamma}$ is related to the radius of the cavity $R$ with $k_{1\gamma}=\omega_{1\gamma}/c=x_{1\gamma}/R$. 

The interaction with the uniform magnetic field $\bm{B}_0$, introduced in Eq.~(\ref{15}), sets the cavity in resonance with the energy level splitting of nanomagnet spin states whenever the relation $\hbar\omega_\gamma = g\mu_B B_0$ is satisfied. Therefore, any spin flip up (down) process of the nanomagnet spins results in an absorption (emission) of a cavity photon in the case of exact resonance, e.g. an applied uniform magnetic field of $B_0 = 7$ T, corresponding to a precession of the macrospin with a frequency of $\sim 200$ GHz, will cause the nanomagnet spins to be in exact resonance with a cavity volume of $1.25$ mm$^3$. We assume the lowest TM mode of the cavity is in resonance with the spin-flip transitions of the nanomagnet, so as higher-energy modes will not be in resonance the subscript $\gamma$ will be omitted from Eq.~(\ref{15}).

The eigenstates of the nanomagnet, treated as a macrospin, are simultaneous eigenstates of the total spin operators $\bm{S}^2$, and $\bm{S}_z$ given by $|l_s,m_s\rangle$, where $|m_s|\leq l_s\leq N/2$. Part of the macrospin approximation is the assumption that $l_s$ is fixed, and most likely it will be the maximal spin state $l_s = N/2$ due to additional energy requirement of any other $l_s\ne N/2$ subspace. The Hilbert space of $N$ independent spins should include the states of a macrospin corresponding to $l_s = N/2$. Therefore, the structure of these basis states is similar to those of the Dicke model\cite{Dicke1954} for $N$ independent atomic spins, wherein $l_s$ is the \textit{cooperation number} of the paramagnetic collection of spins. However, for a realistic nanomagnet, elements of the Hilbert space with $l_s\ne N/2$ are split off in energy due to the exchange interaction giving rise to extra mechanisms, i.e. elementary excitation of spin waves (magnons). 

Since each magnon excitation reduces the total magnetic moment ($\bm{\mu}\propto l_s$) of the nanomagnet in the amount of $2.21\mu_B$ for Fe, it is possible for the nanomagnet total spin angular momentum to start in a different $l_s$ subspace rather than the maximal $l_s=N/2$.  This reduction in $l_s$ is less than $1\%$ at room temperature for iron, suggesting that nanomagnet oscillators of approximately these sizes can be well-described by as having maximal spin at room temperature. The validity of the macrospin approximation relies on the effectiveness of the exchange-locking of the spins at room temperature. For the nanomagnets we consider here, spherical nanomagnets  of radius $r_0 \cong 2.3$~nm, $11$~nm, and $50$~nm consisting of iron (magnetic moment $2.21\mu_B$ per atom), and possessing $N\sim 10^4$, $10^6$, and $10^8$ electron spins, respectively, the macrospin approximation is reasonable\cite{Berkov2008} (although perhaps questionable at for the largest nanomagnet considered).

\section{Properties of the Coupled Nanomagnet-Cavity Hamiltonian}

The total excitation number 2$\xi$, corresponding to the maximum number of photons $n$ in the cavity (when the nanomagnet is parallel to the static magnetic field), needs to be conserved by the Hamiltonian in Eq.~(\ref{5}).
For an initial configuration of the macrospin pointing antiparallel to the static field $\bm{B_0}$ and no photons in the cavity, $\xi=N/2$, the basis states of the spin-photon mode system $|n,m_s\rangle$ can be written as  $|n,\xi-n\rangle$ or $|\xi-m_s,m_s\rangle$, so that the basis states are indexed either solely by photon number of the cavity ($n$), or by eigenvalue of $\bm{S}_z$ ($m_s$).

\begin{figure}[htp]
  \centering
  \includegraphics[width=8.0 cm]{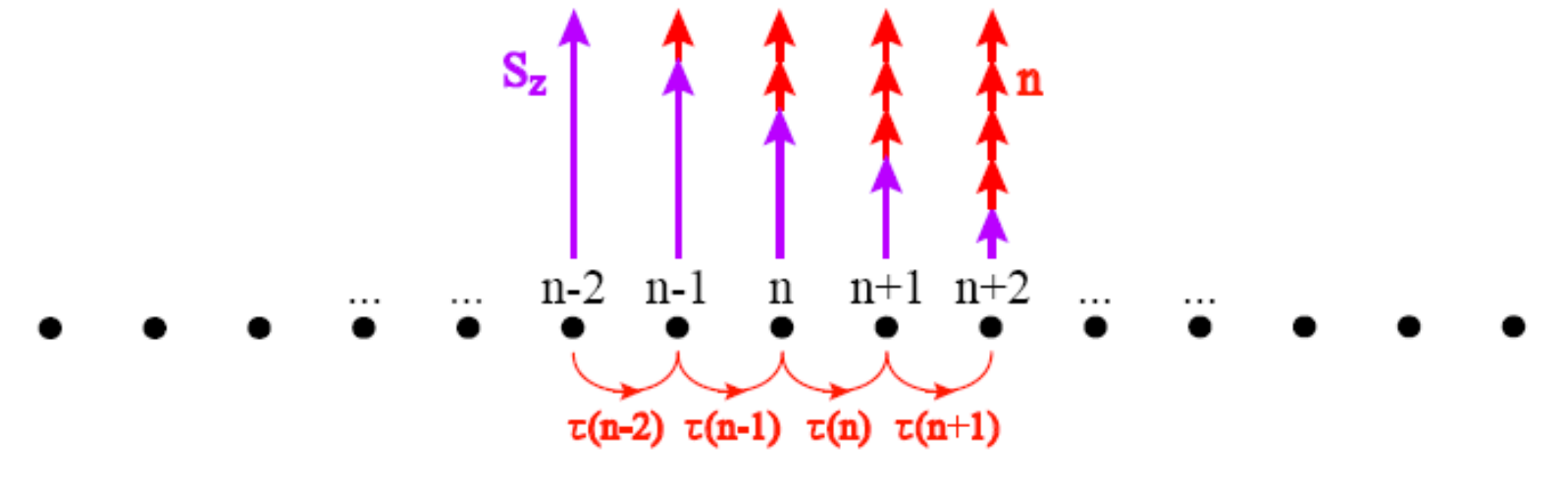}
  \caption{(color online) Lattice-like schematic of the spin-cavity Hamiltonian in Eq.~(\ref{19}) where successive lattice sites represent the possible photon states in the cavity. Note that conservation of total excitation number $\xi$ can be seen from the addition of arrows belonging to the nanomagnet spin states along the z-axis $S_z$ (\textit{purple}, \textit{long arrows}) and the corresponding cavity photon number $n$ (\textit{red}, \textit{short arrows}) for each site. Transitions between successive photon states (lattice sites) are governed by the magnet-microwave mode coupling $\tau(n)$, similar to the hopping in a tight-binding model.}\label{fig:lattice}
\end{figure}

To proceed, we adopt the notation $|n,\xi-n\rangle$ and drop the redundant reference to the $m_s$, so the total Hamiltonian takes the form of
\begin{equation}
\mathscr{H}=\sum_{n=0}^{2\xi}E_0|n\rangle\langle n|-\tau(n)\left[ |n+1\rangle\langle n| + |n\rangle\langle n+1|\right],\label{17}
\end{equation}
in the Fock space, where the constant energy coefficient $E_0$ term and the coupling strength $\tau(n)$ are defined as
\begin{eqnarray}
E_0&=&\hbar\omega\left( \xi+1/2\right),\nonumber\\
\tau(n)&=&\hbar\Gamma g\mu_B(n+1)\sqrt{2\xi-n}\,\, .\label{18}
\end{eqnarray}
In matrix form, the same Hamiltonian can be written as
\begin{equation}
\mathscr{H}=\left(
\begin{array}{ccccc}
E_0 & -\tau(0) & 0 & \cdots & 0 \\
-\tau(0) & E_0 & -\tau(1) & \cdots & 0 \\
0 & -\tau(1) & E_0 & \cdots & 0 \\
\vdots & \vdots & \vdots & & \vdots \\
0 & 0 & \cdots & -\tau(2\xi-1) & E_0
\end{array}
\right),\label{19}
\end{equation}
similar to the Hamiltonian matrix expected for a nearest-neighbor tight-binding model with a spatially-dependent mass (see Fig.~\ref{fig:lattice}). For $2\xi=N=10^8$, the magnet-microwave mode coupling, $\tau(n)$, changes over a range of $33$ kHz - $1.3$ THz through all possible photon (spin) numbers. $\partial \tau(n)/\partial n = \tau^\prime(n)$ acts like a driving force for a fictitious particle moving between sites labeled by photon number $n$, so
$|0\rangle\rightarrow ...\, ...\rightarrow |n-1\rangle \rightarrow |n\rangle \rightarrow |n+1\rangle\rightarrow ...\, ...\rightarrow |2\xi\rangle$. The solutions $n_o$ of $\tau^\prime(n)|_{n_0}=0$ are equilibrium points in cavity photon number, and for this system there is one at $n_0=(4\xi-1)/3$. The coupling can also be expressed in terms of the collective spin number $m_s$ as $\tau(m_s)=\hbar\Gamma g\mu_B(\xi-m_s+1)\sqrt{\xi+m_s}$, with an equilibrium point of $m_0=(1-\xi)/3$. For a system consisting of a very large number of spins ($\xi\gg 1$), the eigenfunctions of the Hamiltonian in Eq.~(\ref{7}) are expected to be centered about $n_0=4\xi/3$ as well as $m_0=-\xi/3$.

For an initial state $|n,m_s\rangle$, if we are only interested in transitions which conserve energy and in which a photon is emitted, the rate of photon emission $R_n$ is proportional to $\sum_{\forall \Psi}|\langle\Psi|a^\dagger S_- |n,m_s\rangle |^2$, where $|\Psi\rangle$ represents the possible final states of the system. Therefore, $R_n=A(n+1)^2(2\xi -n)$, or equivalently $R_n=A(\xi -m_s+1)^2 (\xi+m_s)$. The factor $A$ can be identified as the Einstein A-coefficient by applying $R_n$ to a single spin pointing upward ($\xi=m_s=1/2$) when the cavity has no photons ($n=0$). Since $R_n$ reaches its maximum value of $4A(N/3)^3$ for the equilibrium point $m_0$ (or $n_0$) in the large spin limit, the equilibrium points $n_0$ and $m_0$ are the photon number and spin number, respectively where the nanomagnet-cavity system exhibits \textit{superradiance}\cite{Dicke1954}.

\subsection{Solutions in the Continuum Limit}

For $N=10^4$, $10^6$, and $10^8$ the solutions of the nanomagnet-cavity Hamiltonian corresponds to the diagonalization of large matrices in the form of Eq.~(\ref{19}) with increasing ranks of $10^4$, $10^6$, and $10^8$ for nanomagnets of radius $r_0\simeq 2.3$nm, $11$nm, and $50$nm, respectively. The magnet-photon coupling strengths at the superradiance regime $\tau(n_0)$ are estimated to be roughly $5.3$ neV, $5.3$ $\mu$eV, and $5.3$ meV for these three different nanomagnet sizes.

\begin{figure}[htp]
  \centering
  \includegraphics[width=8 cm]{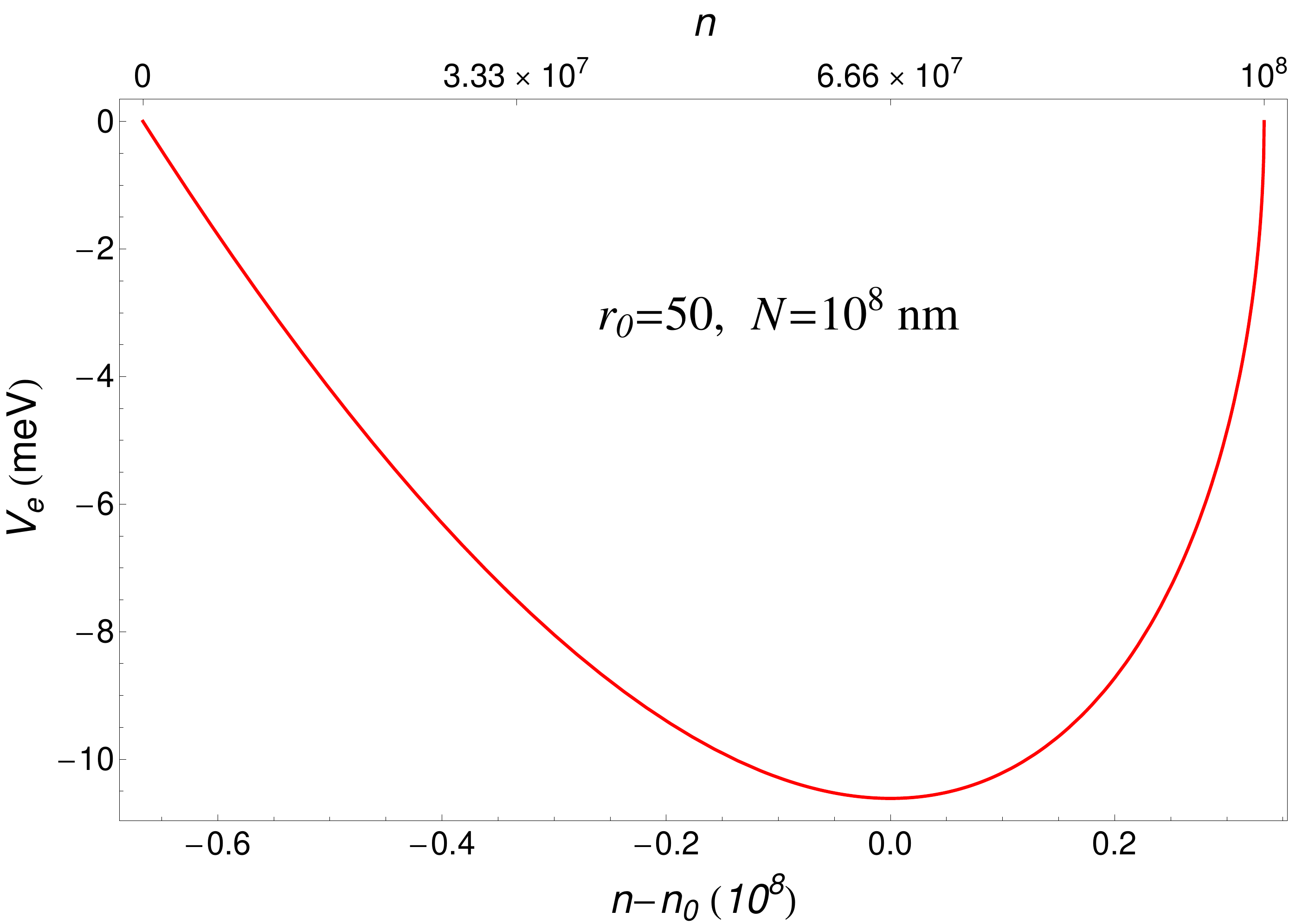}
  \caption{(color online) The effective potential of the magnet-photon system in the WKB approximation is shown with respect to cavity photon number $n$ centered about the superradiance regime $n_0$.}\label{fig:potential}
\end{figure}

\begin{figure}[htp]
  \centering
  \includegraphics[width=8.5 cm]{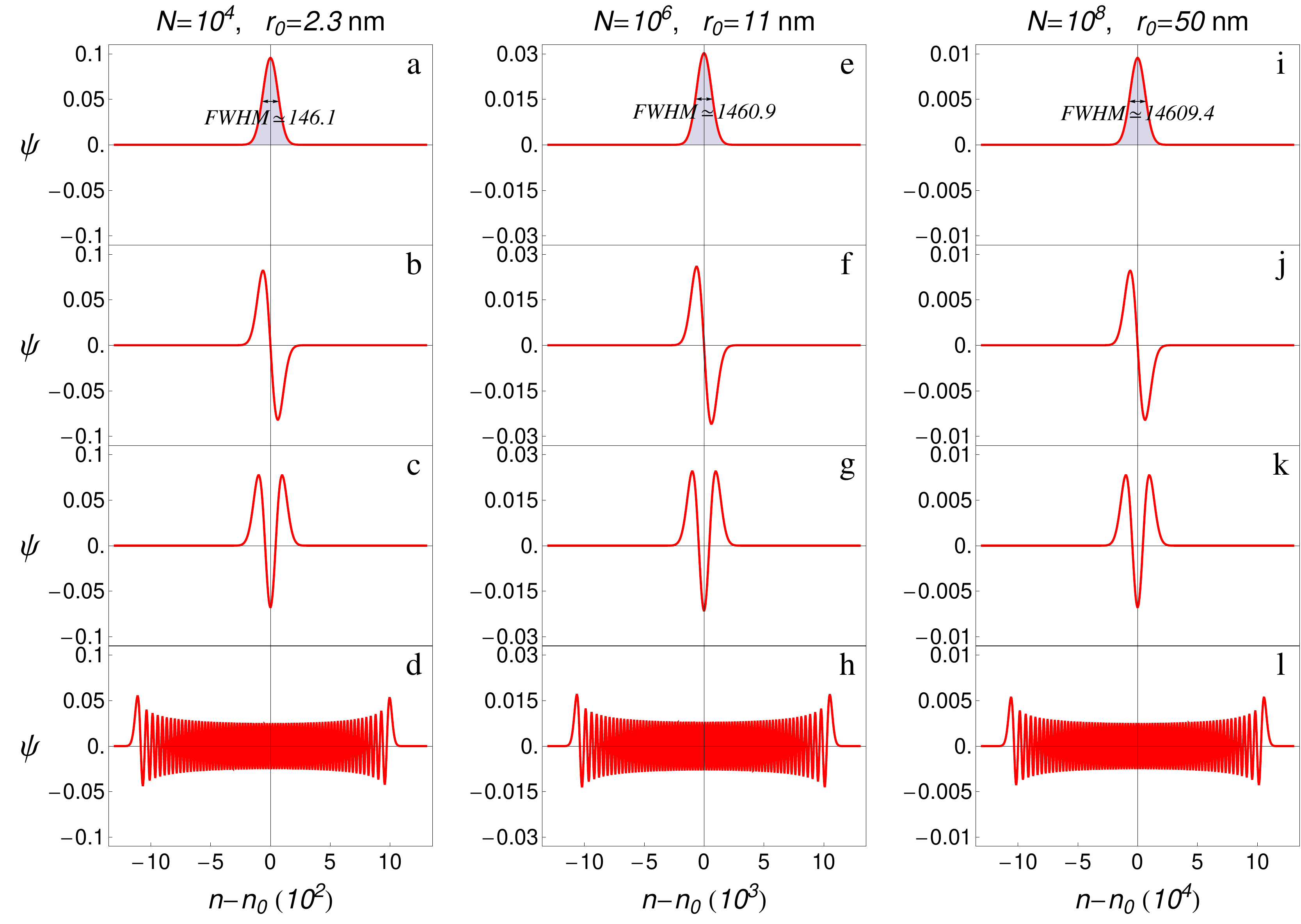}
  \caption{(color online) Wave functions of the nanomagnet-cavity system shown as a function of photon number, $n$, centered about $n_0$, for nanomagnets  of radius $r_0=2.3, 11, 50$ nm, consisting of $N=10^4$, $N=10^6$, and $N=10^8$ spins respectively. First row (a)-(e)-(i) are the ground states with a full width half maximum (FWHM) represented in photon numbers, second row (b)-(f)-(j) are the first excited states, third row (c)-(g)-(k) are the second excited states, and the fourth row (d)-(h)-(l) are the $150^{th}$ excited states.}\label{fig:wavefunctions}
\end{figure}

The eigenfunctions of the nanomagnet-cavity Hamiltonian given in Eq.~(\ref{17}) can be expanded
\begin{equation}
\Psi_j=\sum_{n^\prime}^{2s_z}\psi_j^{n^\prime} |n^\prime\rangle,\label{20}
\end{equation}
in terms of Fock number states and the respective phase constants defined by $\psi_j^{n^\prime}$. Applying the Hamiltonian in Eq.~(\ref{17}) onto these states with the aid of Schr\"{o}dinger equation $\mathscr{H}\Psi_j=E_j\Psi_j$, where $E_j$ are the eigenvalues of the nanomagnet-cavity system, yields the following recursion relation
\begin{equation}
(E_j-E_0)\psi_{j}^{n}+\tau(n-1)\psi_{j}^{n-1}+\tau(n)\psi_{j}^{n+1}=0,\label{21}
\end{equation}
for the phase constants. Since the nanomagnet posseses a very large number of spins, the continuum limit consists in making the replacement $\psi_j^n\rightarrow\psi_j(n\varepsilon)$ for the discrete phase constants in Eq.~(\ref{21}). Then a continuous lattice-like relation can be obtained,
\begin{equation}
E_j \psi_j(n\varepsilon)+\tau(n\varepsilon)\psi_j(n\varepsilon+\varepsilon)+\tau(n\varepsilon-\varepsilon)
\psi_j(n\varepsilon-\varepsilon)=0,\label{22}
\end{equation}
which can also be transformed into the ordinary differential equation
\begin{eqnarray}
&&\tau(x)\frac{d^2\psi_j(x)}{dx^2} + \frac{d\tau(x)}{d x}\frac{d\psi_j(x)}{d x} \label{23}\\
&&+\left( 2\tau(x)-\frac{d\tau(x)}{d x}+\frac{1}{2}\frac{d^2\tau(x)}{d x^2}+E_j\right)\psi_j(x)=0,\nonumber
\end{eqnarray}
with boundary conditions $\psi_j(0)=\psi_j(2s_z)=0$, by Taylor-expanding the phase constants $\psi_j$ in Eq.~(\ref{22}) up to  ${\cal O}(\varepsilon^3)$ and defining $n\varepsilon= x$. Some of the lowest-lying energy eigenvalues $E_j$ and eigenfunctions $\psi_j(x)$ of this differential equation, shown in Fig.~\ref{fig:wavefunctions}, can be obtained in the WKB approximation from
\begin{equation}
S(E_j)=\frac{1}{2\pi}\oint\sqrt{\frac{E_j-V_{e}(x)}{\tau(x)}}dx=j+\frac{1}{2},\label{24}
\end{equation}
where the effective potential is given by $V_e(x)=\tau^{\prime}(x)-{\tau^\prime}^2(x)/4\tau(x)-2\tau(x)$ (see Fig.~\ref{fig:potential}). Shown in Fig.~\ref{fig:wavefunctions} are eigenstates of the coupled nanomagnet-cavity system for three different sizes of nanomagnet. 

\begin{figure}[htp]
\centering
\includegraphics[width=8.0 cm]{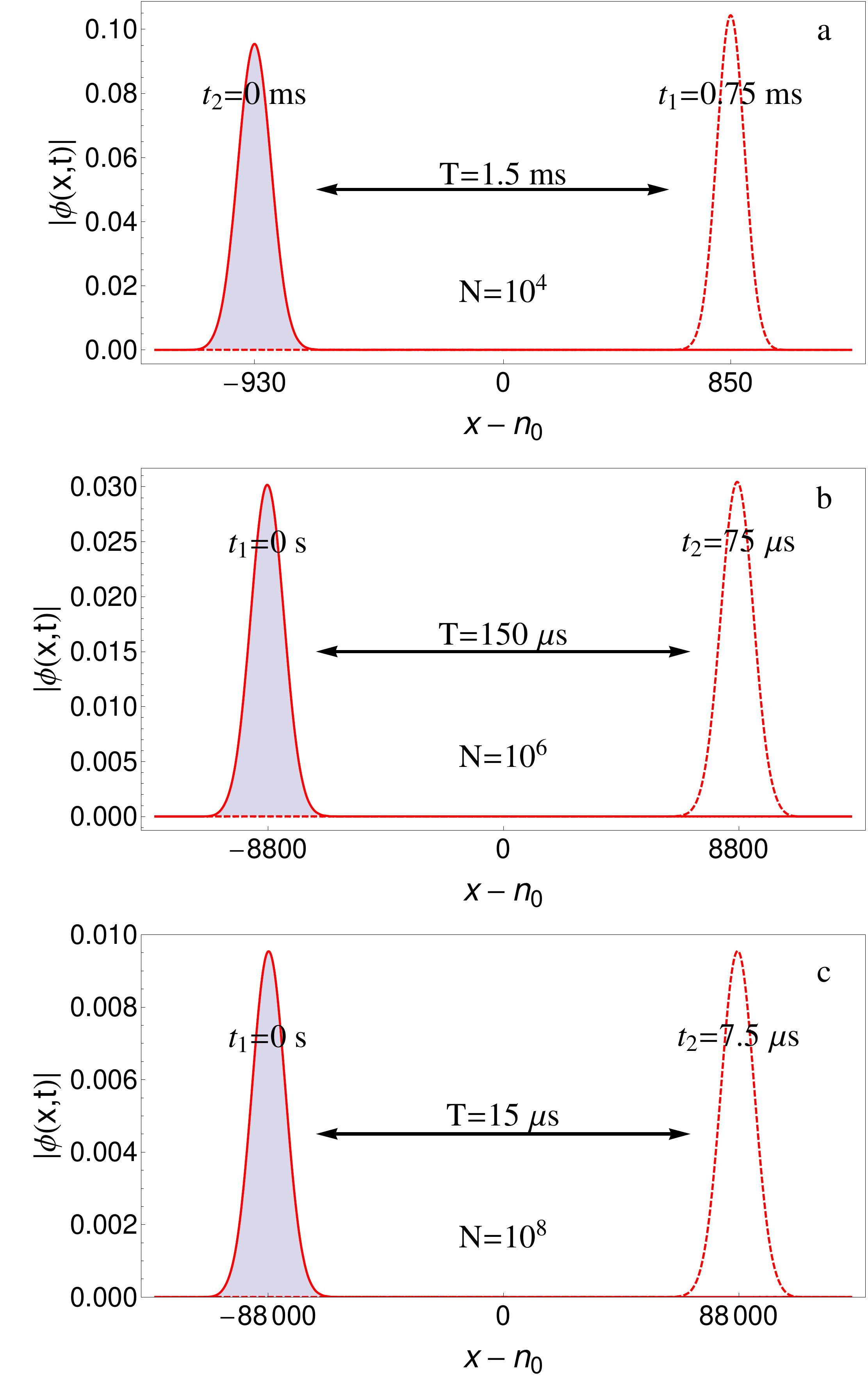}
\caption{(color online) (a) Amplitude of a coherent state for 3 different nanomagnet-photon systems consisting of (a) $N=10^4$, (b) $N=10^6$, and (c) $N=10^8$ spins are shown as a function of photon number $n$. The large oscillations of these coherent states occur about (a) $n_0\sim 6666$ with a period of $T=1.5 ms$, (b) $n_0\sim 6.66\times10^5$ with a period of $T=150 \mu s$, and (c) $n_0\sim 6.66\times10^7$ with a period of $T=15 \mu s$, respectively.}\label{fig:coherent}
\end{figure}
\section{Nanomagnet-Cavity Coherent Dynamics}

\subsection{Form of the Coherent State}

\begin{figure}[htp]
\centering
\includegraphics[width=8.5 cm]{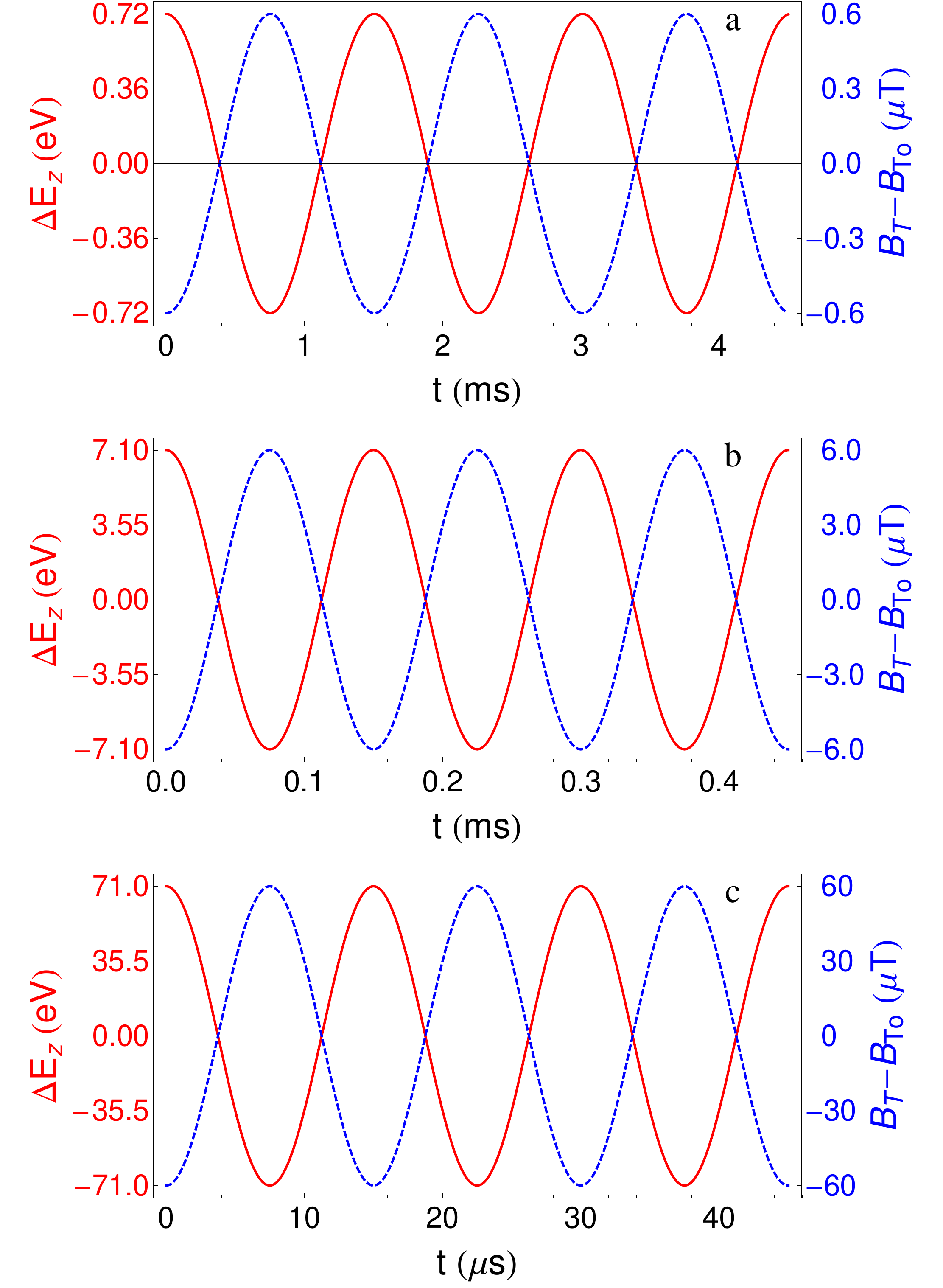}
\caption{(color online) Time evolution of the Zeeman energy of the nanomagnets (\textit{red}, \textit{solid}) consisting of (a) $N=10^4$, (b) $N=10^6$, and (c) $N=10^8$ spins are shown in coherent state representation as well as the amplitude of the transverse magnetic mode of the cavity field (\textit{blue}, \textit{dashed}) at nanomagnet location $z=d$.}\label{fig:oscillation}
\end{figure}

A coherent state for the nanomagnet-cavity system can be written as a displaced nanomagnet-cavity ground state (by a photon number $x_0$ from the equilibrium point $n_0$), or
\begin{equation}
\psi_0(x)=\frac{1}{\sigma\sqrt{2\pi}}e^{-((x-x_0)-n_0)^2 /2\sigma^2}, \label{25}
\end{equation}
where the standard deviation $\sigma$ can be found by matching the values of the full width at half maximum (FWHM) of the ground state and the Gaussian function to each other; for instance ${\rm FWHM}\left[\psi_0(x)\right]=2\sqrt{2\ln 2}\sigma\simeq 14609$ (Fig.~\ref{fig:wavefunctions} (i)) for $N=10^8$. The eigenfunctions of the nanomagnet-cavity system are complete and orthonormal, hence they serve as a suitable basis to expand any coherent state over,
\begin{equation}
\phi(x,t)=\sum_{j=0}^{j_0}A_j e^{-i E_j t/\hbar} \psi_j(x). \label{26}
\end{equation}
Equating the Gaussian function in Eq.~(\ref{25}) to the coherent state of Eq.~(\ref{26}) at initial time $t=0$, i.e.
\begin{equation}
\sum_{j=0}^{\infty}A_j\psi_j(x)=\frac{1}{\sigma\sqrt{2\pi}}e^{-((x-x_0)-n_0)^2 /2\sigma^2},\label{27}
\end{equation}
multiplying both sides by $\psi_j^\prime$ and using the orthonormality condition of the nanomagnet-cavity wavefunctions reveals the phase constants $A_j$ of the expansion as
\begin{equation}
A_j=\frac{1}{\sigma\sqrt{2\pi}}\int_{0}^{2\xi}\psi_{j}(x)e^{-((x-x_0)-n_0)^2 /2\sigma^2}dx.\label{28}
\end{equation}
For three sizes of the nanomagnet, the coherent states shown in Fig.~\ref{fig:coherent}(a)-(c), are characterized by large oscillations over ranges of ($2x_0$=) $1780$, $1.76\times 10^4$, and $1.76\times 10^5$ photons with periods of $T=1.5$~ms, $T=150$~$\mu$s, and $T=15$~$\mu$s, respectively. Summation over the first 150 eigenstates ($j_0=150$) extracted from WKB is sufficient enough to obtain convergence in the dynamical properties of these nanomagnets. The Zeeman energy of the nanomagnet and transverse magnetic field amplitude of the cavity at the nanomagnet's location can also be evaluated from
\begin{eqnarray}
\langle\Delta E_z\rangle&=&\langle\phi(x,t)|\mu_z B_0|\phi(x,t)\rangle,\nonumber\\
\langle B_T\rangle&=& \langle\phi(x,t)|\bm{H}_{TM}(d)|\phi(x,t)\rangle,\label{29}
\end{eqnarray}
respectively, by using the same coherent state representation. Large oscillations of these quantities shown in Fig.~\ref{fig:oscillation} indicates the coherent energy exchange occuring back and forth between photons in the cavity and the spin states of the nanomagnets.

\begin{figure}[htp]
\centering
\includegraphics[width=8.5 cm]{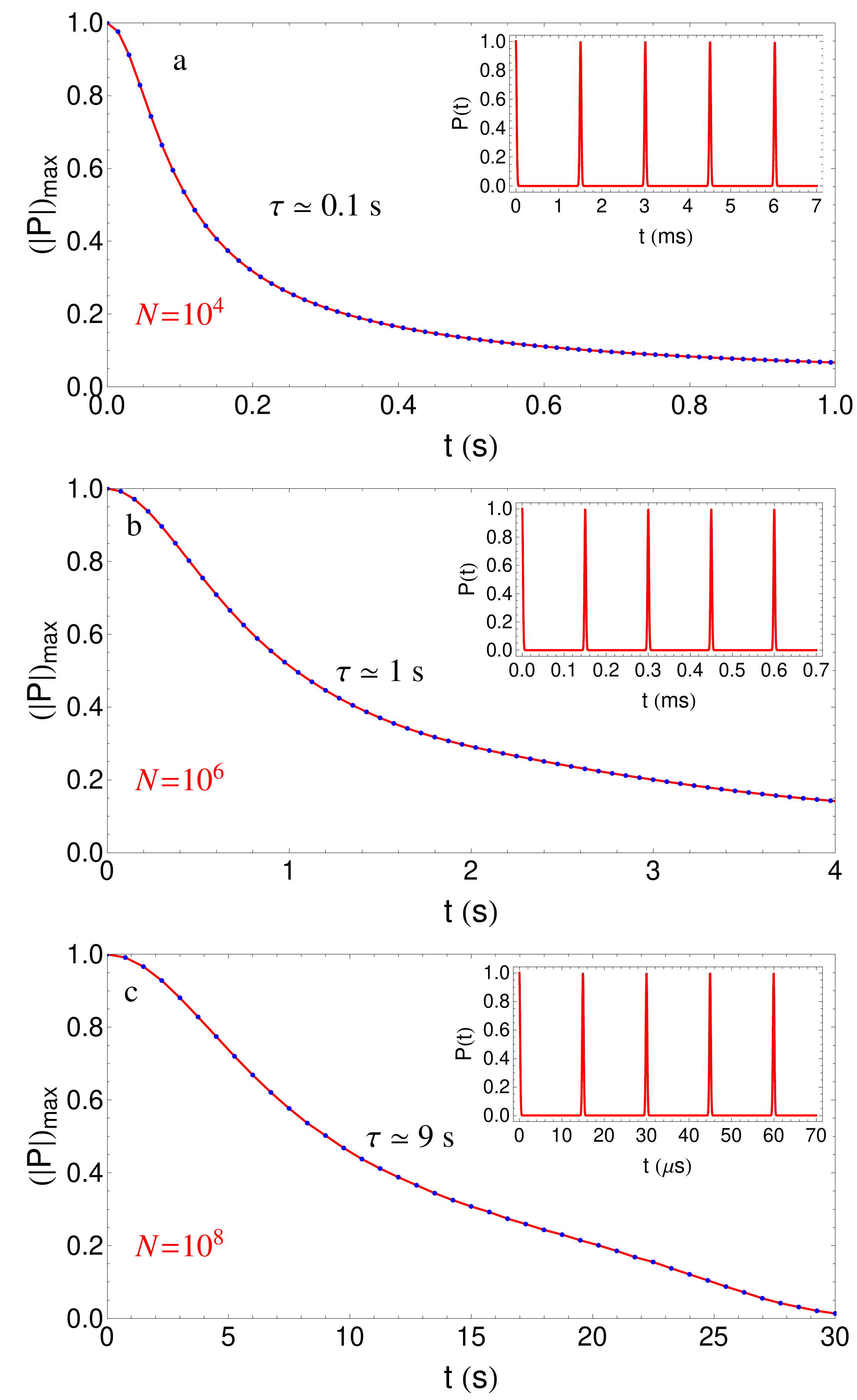}
\caption{(color online) Dephasing time of the coherent state for nanomagnet-photon systems of (a) $N=10^4$, (b) $N=10^6$, and (c) $N=10^8$ spins (or equivalently photons) obtained by a Gaussian fit to the peak values of the dephasing functions (insets) at successive time intervals. Each peak value represents the amount of correlation after every full period $T$ of oscillation.}\label{fig:dephasing}
\end{figure}

\subsection{Dephasing of the Coherent State}

The coherent properties of this nanomagnet-photon system will also depend on the dephasing of the coherent state $\phi(x,t)$, due to inhomogenity of the coupling $\tau(n)$ in Eq.~(\ref{18}). The dephasing time of the nanomagnet-cavity coherent state can be extracted by a Gaussian fit to the peak values of the autocorrelation function between a coherent state at time $t$ and its initial state at $t=0$,
\begin{eqnarray}
P(t)&=&|\langle\phi(x,t)|\phi(x,0)\rangle|^2,\label{30}\\
&=&\left|\sum_{j=0}^{\infty}|A_j|^2 e^{iE_jt/\hbar}\right|^2,\nonumber
\end{eqnarray}
 whereas each peak (inset of Fig.~\ref{fig:dephasing}) is representing the revival amount of the coherent state after every successful period $T$ of oscillation. Exceptionally long dephasing time of order seconds are  shown in Fig.~\ref{fig:dephasing}.  As the nanomagnet gets bigger the change in $\tau(n)$ with $n$ becomes smoother and smoother, leading to longer dephasing times. 
 
 Although this treatment is for zero temperature, the coherent properties of the nanomagnet-photon system should persist to as high a temperature (and over as long a timescale) as the macrospin description remains reliable. We have assumed an infinite $Q$ for the cavity, so the decoherence of the system is expected to be determined by photon leakage from the cavity, rather than these exceptionally long calculated times. Furthermore, the elementary spin excitations (magnons) would not directly affect the dephasing of the system, for magnons preserve the spin quantum number $m_s$, requiring an up spin to flip down for every down spin flipping up. In realistic nanomagnets, spin-lattice coupling of $m_s$ to phonons through spin-orbit coupling will cause a cutoff of the dephasing times shown in Fig.~\ref{fig:dephasing}. For spheres of yttrium iron garnet (YIG) at low temperature this spin-lattice time is several $~\mu$s. \cite{LeCraw1962,Sparks1960} Therefore, observation of a full oscillation cycle should be possible for nanomagnets with a radius of $50$~nm or larger. On the other hand, the times at room temperature in YIG ($\sim 200$)~ns\cite{LeCraw1962} and iron ($\sim 20$~ns)\cite{Frait1980} are too small to observe a full oscillation. However, coherent dynamics corresponding to a portion of the oscillation involving $\sim 24$~photons/ns, or $\sim 470$~photons for iron and $4700$ photons for YIG should be still observable for the nanomagnet with radius $r_0=50$~nm. If, however, the modal coupling is increased using approaches such as tip-enhancement of the optical field, then the coupling could be far stronger even for a small nanomagnet. Guided by estimates from tip-enhanced Raman spectroscopy\cite{Moskovits1985}, the intensity of the mode at the nanomagnet's position could be increased by $10^2-10^6$, leading to enhancements of the oscillation frequency of order $10-10^3$. 

\subsection{Crystalline Magnetic Anisotropy}

We also examine other deviations from ideality for the nanomagnet, such as the spin dependent cubic crystalline magnetic anisotropy (CMA). The CMA of iron is given by
\begin{equation}
E_{CMA}=U_1\left(\kappa_1^2\kappa_2^2+\kappa_2^2\kappa_3^2 +\kappa_1^2\kappa_3^2\right) +U_2\kappa_1^2\kappa_2^2\kappa_3^2,\label{31}
\end{equation}
where $U_1=4.2\times 10^5$ erg/cm$^3$ and $U_2=1.5\times 10^5$ erg/cm$^3$ are the cubic anisotropy constants for iron at room temperature and an arbitrary magnetization direction is defined by the directional cosines $\kappa_1$,$\kappa_2$,$\kappa_3$ referred to the cube edges. Since the nanomagnet is a sphere, shape anisotropy is not relevant. In the case of a cubic crystal whose easy axis is aligned along the body diagonal, $E_{CMA}$ energy depends on the orientation of the nanomagnet spin $\bm{S}$, defined by $\kappa_i$. 

The CMA of iron causes a detuning of the energy spacing for different spin orientations from the resonant frequency of the cavity, along with a dispersion in that spacing. The uniform detuning, corresponding to  a uniform shift in the precession frequency of the nanomagnet, can be compensated for with a slight adjustment in the applied magnetic field. The dispersion, however causes a variable detuning of  roughly $200$ neV, $13$ neV, and $1.3$ neV of the $E_0$ in Eq.~(\ref{17}) over the range of oscillation shown in Fig.~\ref{fig:coherent}(a)-(c), respectively. For the smallest nanomagnets the effect of CMA dominates over the coupling between the photons and the spin. For example, for a nanomagnet radius of $2$ nm consisting of $10^4$ total spins, the CMA is significantly larger than the magnet-photon coupling strength $\tau(n)$ ($\sim 5.3$ neV) in Eq.~(\ref{17}). Therefore the CMA will cause the eigenstates to localize in photon and spin number, producing rapid decoherence for a coherent state. We note that this observation largely rules out the possibility of observing these coherent oscillations in a single molecular magnet\cite{Bogani2008}, for the spins of these molecules are considerably smaller than the spin of the nanomagnet considered above.  However, this detuning is much smaller than the magnet-photon coupling strength of other nanomagnet sizes (10 nm and 50 nm in radii) and therefore will not destroy the coherent oscillations for them, although it may still limit the dephasing times to shorter than that shown in Fig.~\ref{fig:dephasing}(b)-(c).

\section{Concluding Remarks}

Calculations for three different nanomagnet sizes in a photonic cavity indicate that strong-field coupling between photons and spins is possible, and should substantially exceed the coupling observed in solids between orbital transitions and light. The Hamiltonian for the coupled nanomagnet-cavity system is solved in the continuum limit to obtain a coherent state representation of the system around the superradiance regime. This coherent state is characterized by large oscillations in photon number of the cavity (or equivalently the total spin number of the nanomagnet) with exceptionally long dephasing times and is expected to be observable for realistic nanomagnets with radii from $10-50$~nm. Approaches to enhance the coupling, such as using a metal tip to enhance the optical field, have been proposed. For the smallest nanomagnet (2~nm radius) the dispersion caused by crystalline magnetic anisotropy would largely quench the coherent oscillations, but for nanomagnets in the $10-50$~nm radius range the coupling to the cavity is much stronger than the dispersion caused by CMA.  The dephasing times increase with increasing nanomagnet size, due to the greater uniformity of the coupling terms between states that differ by one photon and one spin flip. Thus the most coherent nanomagnet-cavity systems will be those that are just under the size threshold where the macrospin approximation ceases to be accurate.  The effects of magnons have been considered and shown to not substantially modify these results.

Future work shall investigate how to use the strong coupling features described here to transfer coherently states of the electronic system to the photonic one, and back again. A particularly interesting direction will be to consider the effect of active nanomagnetic systems, such as those demonstrated  to be coherently driven by electrical spin currents\cite{Berger,Slonczewski,Buhrman,Kiselev2003,Sankey2006,Urazhdin2005}, on the optical state of the cavity. As phase-locking has been demonstrated between two such oscillators\cite{Kaka2005,Mancoff2005}, mediated perhaps by spin waves, would it be possible to phase-lock them through interaction with a cavity such as the one considered here?

\begin{acknowledgements}
We thank A. Kent and D. C. Ralph for helpful discussions. This work was supported by an ONR MURI. 
\end{acknowledgements}

\end{document}